\documentclass[12pt,]{article}
\usepackage{amsfonts}
\usepackage{amsmath,amssymb,bm}
\usepackage{graphicx}

\input{tcilatex}

\begin{document}

\title{A note on Gribov copies in 3D Chern-Simons theory}
\author{Fabrizio Canfora \\
%EndAName
{\small \textit{Centro de Estudios Cientificos (CECS), Casilla 1469
Valdivia, Chile.}}\\
{\small e-mails: \textit{canfora@cecs.cl}}}
\date{}
\maketitle

\begin{abstract}
Using powerful tools of harmonic maps and integrable systems, all the Gribov
copies in the Coulomb gauge in 3D Chern-Simons theory are constructed. Some
issues about the Gribov and the modular regions are shortly discussed. The
Gribov copies of the vacuum in 3D QCD in the Coulomb gauge are described. An
interesting implication of the presence of Gribov copies is briefly pointed
out.
\end{abstract}

%\maketitle


Preprint: CECS-PHY-08/02

\section{Introduction}

The Gribov ambiguity \cite{Gri78} is one of the most deep non perturbative
phenomena in gauge theories: it is a global obstruction in achieving a
global gauge fixing in linear derivative gauges (like Lorentz, Coulomb,
Landau and so on) related to the non trivial topology of the space of
non-Abelian gauge connections\footnote{%
An interesting ambiguity, which in a sense is dual to the Gribov's one, was
discovered in \cite{Wil}: the authors showed that there may exist gauge
potentials which are not gauge equivalent but which generate the same
curvature.}. Furthermore, as Gribov himself pointed out, it seems that such
an ambiguity is closely related to the confinement in QCD once the path
integral is restricted to a ambiguity-free region (see, \cite{Zwa96} and
references therein; for two detailed reviews see \cite{SS05} \cite{EPZ04}).
Recently, it appeared an interesting paper \cite{ILM07} in which an
alternative argument has been presented which suggests a close relation
between Gribov ambiguity and confinement\footnote{%
At a first glance, one could simply use other gauge fixings free of
ambiguities in order to disproof such a relation but in four dimensions very
often such gauge fixings have their own problems (see \cite{EPZ04} and
references therein).}. Such an argument is intriguing in that it seems not
to depend heavily on the explicit form of the action so that it could be
applied to other contexts.

This would provide some recent results in gravity (see, for instance, \cite%
{Re98} \cite{MS07}, \cite{Li06}, \cite{Per06}, \cite{Li07} and references
therein) with an interesting physical interpretation. In such works,
following an idea first pointed out by S. Weinberg in \cite{We79} (see, for
a related work on the same line, \cite{GW96}), the authors found strong
evidence of an UV fixed point for gravity which could overcome the
perturbative non-renormalizability of gravity \cite{GS86}. Such an UV fixed
point could correspond to a confinement phase transition in which only
scalar fields survive in the physical spectrum improving the UV behavior of
gravity as first pointed out in \cite{Ca05} \cite{Ca06} (see also \cite{To77}
\cite{Smo82}).

However, many point about the relations between Gribov ambiguity and
confinement have to be clarified. In particular, the Gribov ambiguity is
also present in field theories with a vanishing beta function such as N=4
SUSY Yang-Mills in four dimensions and Chern-Simons theory in three
dimensions (such a theory was introduced in the physical literature in the
seminal paper \cite{DeJaTe}). It is therefore interesting to analyze the
issue of Gribov copies in Chern-Simons case (which is simpler than the case
of N=4 SUSY Yang-Mills but highly non trivial) to see how such an ambiguity
manifests itself in this simpler case. Furthermore, there is no common
agreement in the literature about whether or not one has to restrict the
path integral to an ambiguity free region. In particular, in \cite{FL96} it
has been constructed a solvable model (whose BRS analysis has been provided
in \cite{Fuj96}) in which one has to sum over all the copies instead of
restricting to an ambiguity free region.

The analysis of Gribov ambiguity in 3D Chern-Simons theory can help to shed
light on the above interesting questions in a context which is far simpler
than 4D QCD.

The paper is organized as follows: in the second section the Gribov
ambiguity in the Chern-Simons case is shortly analyzed. In the third
section, the Gribov equation and its solutions are presented. In the fourth
some possible implications of the presence of Gribov copies in Chern-Simons
theory are pointed out. Finally, conclusions and perspectives are presented.

\section{The Gribov ambiguity in Chern-Simons}

The Gribov problem was discovered studying the Faddev-Popov procedure for
quantizing Yang-Mills theory using path integral methods \cite{Gri78}. To
carry on the Faddev-Popov procedure it is necessary to choose a gauge
fixing. Usually, for practical computations, the more convenient gauge
fixings are the Coulomb, the Landau, the Lorentz and so on. However, such
gauge fixings do not fix the gauge in the non Abelian case: there are gauge
equivalent connections fulfilling the same (Coulomb, the Landau, the Lorentz
and so on) gauge conditions. The Gribov ambiguity is not a special feature
of path integral quantization: it also appears in the canonical formalism
when applying the Dirac procedure (see, for instance, \cite{HT}). In gauge
systems with first class constraints one needs to introduce suitable gauge
fixing functions in such a way that these gauge fixing functions together
with the first class constraints form a second class system of constraints
to be analyzed with the Dirac method. In the presence of Gribov ambiguities
such a procedure only works locally.

Let us see in detail the 3D Chern-Simons case: the action is%
\begin{equation}
S=k_{bare}S_{0}+S_{gf}\left( c,\overline{c},A,\lambda \right) ,\ \ S_{0}=%
\frac{1}{4\pi }\int_{M}tr\left( AdA+\frac{2}{3}A^{3}\right)  \label{cs1}
\end{equation}%
where the connection $A$ takes values in the algebra of $SU(N)$, the bare
coupling constant $k_{bare}$ is an integer, the trace is in the fundamental
representation, $M$ is a smooth three dimensional manifold. $S_{gf}$
represents the gauge fixing and the ghosts terms\footnote{$\lambda $ is the
Lagrange multiplier enforcing the gauge fixing, $c$ and $\overline{c}$\ are
the ghost and the anti-ghost and in the non-Abelian case the Faddev-Popov
determinant also depends on $A$.} which break the diffeomorphisms invariance
of the first term due to the introduction of a background metric (which will
be assumed to be flat and of Lorentzian signature). Here the Coulomb gauge
(which allows to use powerful results in the theory of harmonic maps and
integrable systems) will be considered%
\begin{equation}
\chi =\partial _{i}A^{i}=0  \label{cgf1}
\end{equation}%
in which the splitting between time ($0\sim t$) and space ($i\sim x,y$)
indices will be introduced in a moment\footnote{%
An interesting discussion of the possible phenomenological relevance of
Gribov copies in three dimensional gauge theories can be found in \cite{T98}.%
}. To the best of author's knowledge, the first application of harmonic map
theory to find Gribov copies in the four dimensional $SU(2)$\ Yang-Mills
case can be found in \cite{V78}. The Singer theorem \cite{Si78}\ tells that
if Gribov ambiguities are present in the Coulomb gauge they are present in
all the derivative gauges. Let us consider the case in which $M=\Sigma
\times R$ where $R$ will be interpreted as the time direction so that one
can decompose (using the notation of \cite{EMS89}) $d$ and $A$ as follows%
\begin{equation*}
d=dt\otimes \frac{\partial }{\partial t}+\overrightarrow{d},\ \ A=A_{0}+%
\overrightarrow{A},\ \ \overrightarrow{F}=\overrightarrow{d}\overrightarrow{A%
}+\overrightarrow{A}^{2}.
\end{equation*}%
$S_{0}$ now reads%
\begin{equation*}
S_{0}=-\frac{1}{4\pi }\int_{M}tr\left( \overrightarrow{A}\frac{\partial }{%
\partial t}\overrightarrow{A}\right) +\frac{1}{2\pi }\int_{M}tr\left[ A_{0}%
\overrightarrow{F}\right]
\end{equation*}%
in which $A_{0}$ is a Lagrange multiplier enforcing $\overrightarrow{F}=0$
so that $\overrightarrow{A}$ is locally flat. In the path integral formalism
this implies the presence of a delta of $\overrightarrow{F}$: $\delta \left( 
\overrightarrow{F}\right) $. One can formally solve such a constraint as
follows%
\begin{equation*}
\overrightarrow{A}=U^{-1}\overrightarrow{d}U
\end{equation*}%
(where $U$ is a single-valued map from $\Sigma \times R$ to $SU(N)$, we will
be more precise on the choice of the space $\Sigma $ in a moment) so that
the Coulomb gauge condition (\ref{cgf1}) becomes%
\begin{equation}
\partial _{i}\left( U^{-1}\partial ^{i}U\right) =0.  \label{cgf2}
\end{equation}

\section{Gribov equation}

Can one find different solutions of Eq. (\ref{cgf2}) giving rise to gauge
equivalent gauge potentials? To answer to this question in a clear way it is
useful to consider the case in which the space $\Sigma $ is a two sphere 
\begin{equation*}
\Sigma =S^{2}
\end{equation*}%
since in this case the moduli space of flat connections is trivial. If Eq. (%
\ref{cgf2}) would only have the trivial solution $U=U_{0}$ a constant matrix
in the group then there would be no Gribov copies. On the other hand,
non-trivial Gribov copies correspond to different solutions of Eq. (\ref%
{cgf2}) giving rise to gauge equivalent potentials fulfilling the following
boundary conditions:%
\begin{equation}
\int_{\Sigma }tr\left( A_{i}\right) ^{2}d^{2}\Sigma \approx \int_{\Sigma
}tr\left( U^{-1}\partial _{i}U\right) ^{2}d^{2}\Sigma <\infty .  \label{bc1}
\end{equation}%
Some beautiful results in the theory of harmonic maps (see \cite{Ul89}) and
integrable systems (see for instance \cite{Wo89}, \cite{Wa90}) provide one
with the complete classification of the solutions of Eq. (\ref{cgf2})
fulfilling the boundary conditions (\ref{bc1}) for a generic $SU(N)$. Thus,
remarkably enough, one can construct \textit{all the Gribov copies} in the
Coulomb gauge in 3D Chern-Simons theory (in the case in which $\Sigma $ is a
two sphere). In \cite{Ul89} it has been shown that all the solutions of Eq. (%
\ref{cgf2}) fulfilling the condition in Eq. (\ref{bc1}) are of the form%
\begin{equation}
U=U_{0}\dprod\limits_{i=1}^{u}\left( 1-2R_{i}\right) ,\ \ R_{i}^{2}=R_{i}
\label{mus1}
\end{equation}%
where the positive integer $u$ is the so-called \textit{unitons number}, $%
U_{0}$ is a constant matrix in the group and $R_{i}$ are projectors which
satisfy some first order differential equations. The integer $u$ is smaller
than $N$\footnote{%
To the best of the authors knowledge there is no theorem telling that $u=N-1$
but in many cases this is actually so.}: starting, for instance, with the
trivial solution $U_{0}$ one can "dress it" (see \cite{Wo89}, \cite{Wa90}
and references therein) with the factors $1-2R_{i}$ getting new solutions
but this can only be done a finite number of times\footnote{%
This is quite different from the standard dressing techniques in solitons
theory and integrable systems which allow to construct solutions with an
arbitrary large number of elementary solitons.} denoted by $u$. Another
remarkable feature of the Gribov copies in the Coulomb gauge in 3D
Chern-Simons case is that they are labelled by the discrete index.

This is quite different from the QCD case in four dimensions in which the
Gribov equation admits continuous families of solutions: the Coulomb gauge
condition in 3D Chern-Simons gauge theory together with the physical
boundary condition (\ref{bc1}) reduce to the integrable equation of the
harmonic maps. QCD in three dimensions also has continuous families of
Gribov copies due to the fact that, unlike the Chern-Simons case, the space
of classical solutions \textit{is not} made of pure gauge fields. However,
the copies of the vacuum in 3D QCD on $S^{2}\times R$ in the Coulomb gauge
have exactly the same structure as the Gribov copies in 3D Chern-Simons on $%
S^{2}\times R$ since the Gribov equation for the copies of the vacuum
together with the physical boundary conditions reduce to Eqs. (\ref{cgf2})
and (\ref{bc1}). Thus, harmonic map theory allows to construct all the
Gribov copies of the vacuum in 3D QCD on $S^{2}\times R$ in the Coulomb
gauge which are also labelled by the unitons number. This is a peculiar
property of Yang-Mills theory in 3 dimensions: Gribov copies of the vacuum
can be avoided in 4D QCD by asking suitable boundary conditions (see for
instance \cite{SS05}). The complex structures present in a two dimensional
space ($S^{2}$ in the present case) allow the construction of genuine Gribov
copies of the vacuum in the Coulomb gauge in 3D QCD on $S^{2}\times R$.

Thus, in the $SU(2)$ case the more general solutions are 1-uniton solutions.
In the $SU(2)$ case the most general Gribov copy is 
\begin{eqnarray*}
\overrightarrow{A} &=&U^{-1}\overrightarrow{d}U \\
U &=&U_{0}\left( 1-2R\right) \\
\left( 1-R\right) \partial _{+}R &=&0,\ \ R^{2}=R
\end{eqnarray*}%
where 
\begin{equation*}
\partial _{+}=\frac{1}{2}\left( \partial _{x}+i\partial _{y}\right)
\end{equation*}%
and $R$ is a holomorphic projector.

\subsection{Modular vs Gribov regions in 3D Chern-Simons}

An important issue in the analysis of Gribov ambiguity is the relation
between the modular and the Gribov region (see \cite{vb}, \cite{z2}, \cite{s}%
, \cite{z3}). In the present case, the \textit{Gribov region} $\Omega $\ is
defined as the region of local minima of the functional $S_{g}[U]$ of $U$
which is defined on the gauge orbit of $A_{i}$ 
\begin{equation*}
S_{A_{i}}[U]=\int_{S^{2}}dS^{2}tr\left( A_{i}^{U}\right) ^{2}
\end{equation*}%
where $dS^{2}$\ is the volume element on the sphere and $A_{i}^{U}$\ is the
gauge transformed of $A_{i}$ with the gauge transformation generated by $U$
(which, in general, may also be homotopically non-trivial). At the \textit{%
Gribov horizon} $\partial \Omega $ the lowest eigenvalue of the Faddev-Popov
operator vanishes. The \textit{modular region} $\Lambda $\ is defined as the
global minima of $S_{g}[U]$. It can be shown that at the global minima of
the above action, the vector potential is transverse and the Faddev-Popov
operator is positive. The restriction to the interior of the Gribov region $%
\Omega $ is not enough to avoid copies so that, if one wants to restrict the
path integral to an ambiguity free region, then it is necessary to consider
the interior of the modular region $\Lambda $ (since, unless suitable
identifications are performed, copies may appear on the boundary $\partial
\Lambda $). It may also happen that portions of the boundaries of the
modular region $\partial \Lambda $ and of the Gribov region $\partial \Omega 
$ coincide (see \cite{vb}, \cite{z2}, \cite{s}, \cite{z3}). At a practical
level, it is much more difficult to work with the restriction to the modular
region than it is with the restriction to the Gribov region (a nice review
on the implementation of such a restriction can be found in \cite{vb1}).
Being the moduli space of flat connection of $S^{2}$ trivial, the above
action is proportional to the positive definite action which defines the
harmonic maps%
\begin{equation}
S[U]=\int_{S^{2}}dS^{2}tr\left( U^{-1}\partial _{i}U\right) ^{2}
\label{griac2}
\end{equation}%
It is quite obvious that the global minima are the constant solutions $%
U=U_{0}$ which give rise to a vanishing gauge potential so that the interior
of the modular region in this case reduces to the trivial gauge potential.
Thus, one can answer the question of whether or not in this case $\Lambda $
and $\Omega $ coincide by investigating if the non-trivial unitons solutions
are local minima or simply saddle points of the action (\ref{griac2}). In
the $SU(2)$ case the above action is equivalent to the two dimensional $%
%TCIMACRO{\U{2102} }%
%BeginExpansion
\mathbb{C}
%EndExpansion
P^{1}$ model (the generic $%
%TCIMACRO{\U{2102} }%
%BeginExpansion
\mathbb{C}
%EndExpansion
P^{n-1}$ models were introduced in the physical literature in \cite{Ei78} 
\cite{CJ78}, see also \cite{Pol76}) and the more general 1-uniton solutions
are nothing but instanton solutions of the $%
%TCIMACRO{\U{2102} }%
%BeginExpansion
\mathbb{C}
%EndExpansion
P^{1}$ model (to the best of author's knowledge, instantons in the generic $%
%TCIMACRO{\U{2102} }%
%BeginExpansion
\mathbb{C}
%EndExpansion
P^{n-1}$ models were firstly constructed in \cite{DDL78}). If one neglects
the zero modes which correspond to translations of the position of the
instanton and to constant rescalings of its size, well known topological
arguments would suggest that instanton solutions are local minima.

This is a crucial point, to define functional determinants in quantum field
theory it is a standard procedure to "quotient out" trivial zero modes but
in the present case one is asking "are the instanton solutions of the two
dimensional $%
%TCIMACRO{\U{2102} }%
%BeginExpansion
\mathbb{C}
%EndExpansion
P^{1}$ model local minima?" Strictly speaking, the answer is "no" because of
the flat directions: if one change the position (or the size) of the
instanton the value of the action does not change. Consequently, the
Faddev-Popov determinant evaluated at the instanton solutions vanishes and
they are neither in the interior of $\Lambda $ nor in the interior of $%
\Omega $. On the other hand, if one takes the opposite view that instanton
solutions which are related by translations of their positions and/or by
constant rescalings of their size represent the same solution then one has
to factor out the zero modes obtaining a local minimum. In any case, it
should be noted that the instanton solutions of the $%
%TCIMACRO{\U{2102} }%
%BeginExpansion
\mathbb{C}
%EndExpansion
P^{1}$ model are homotopically non-trivial so that the $SU(2)$ Gribov copy
of the vacuum $A_{i}=\left( U_{I}\right) ^{-1}\partial _{i}\left(
U_{I}\right) $\ (where $U_{I}$ is the instanton solution of the $%
%TCIMACRO{\U{2102} }%
%BeginExpansion
\mathbb{C}
%EndExpansion
P^{1}$\ model) is obtained by acting on the vacuum with a homotopically
non-trivial gauge transformation. Thus, one can argue that the Faddev-Popov
determinant vanishes using the elegant argument in \cite{JMR}.

In the generic $SU(N)$ case the situation is less clear: strictly speaking,
the unitons number is not well defined since if one adds, for instance, a
uniton factor to a two-unitons solution the resulting expression may be
equivalent to a 1-uniton solution. For this reason, in \cite{Ul89} the 
\textit{minimal unitons number} was defined as the minimal number of unitons
that are required to construct a given solution of Eq. (\ref{cgf2})
fulfilling the boundary conditions (\ref{bc1}). In the generic case one may
expect that a solution of Eq. (\ref{cgf2}) is simply a saddle point of the
action (\ref{griac2}). Solutions representing local minima should be
characterized by some notion of "minimality" or "indecomposability"
otherwise intuitive physical arguments based on the action (\ref{griac2})
would suggest the possibility that such a solution may "decay" into some
more fundamental solutions but already in the $SU(3)$ case the computations
are quite involved and it is not possible to obtain explicit expressions for
the solutions in the general case. The $SU(2)$ example suggests that in the
generic $SU(N)$ case also there could be instanton-like solutions but, being
homotopically non-trivial, the argument in \cite{JMR}\ ensures the vanishing
of the Faddev-Popov determinant (unless zero modes are quotiented out): this
interesting point deserves further investigation.

\section{Chern-Simons shift}

In Chern-Simons theory there is a well known debate in the literature: the
seminal paper of Witten \cite{Wit89} shed light on the close relations
between knot theory, conformal field theory and Chern-Simons theory. It
allowed the computations of the Jones polynomials in knot theory using
Chern-Simons perturbations theory and some inputs from conformal field
theory. Chern-Simons theory is renormalizable by power counting; the beta
function and the anomalous dimensions of the elementary fields are vanishing
to all orders in perturbation theory (see \cite{DPL90}). The only free
parameter of the model is the renormalized coupling constant $k_{ren}$ which
is fixed by the normalization conditions. Therefore, on the Chern-Simons
side, such polynomials turn out to depend on a parameter $q$ which is
related to the renormalized Chern-Simons coupling constant $k_{ren}$: 
\begin{equation}
q=\exp \left[ -i\frac{2\pi }{k_{ren}}\right]   \label{ske1}
\end{equation}%
and this conclusion should not depend on the regularization scheme. The
debate in the literature is about the relation between $k_{ren}$ and the
bare coupling constant $k_{bare}$ appearing in the action (\ref{cs1}). It
has been shown in \cite{EMS89}\ that the essential point behind the shift%
\begin{equation}
k\rightarrow k+N  \label{ske2}
\end{equation}%
is that in the evaluation of expectation values one must integrate out the
gauge degrees of freedom and in so doing non-trivial Jacobians appear. The
source of the shift is the anomaly in these Jacobians\footnote{%
Indeed, there are regularization schemes which give rise to similar results
in covariant gauges (see, for instance, \cite{ALR90}). However such kind of
results depend on the regularization scheme and with different
regularization schemes one can also obtain shift different from Eq. (\ref%
{ske2}). In any case, it is fair to say that the regularization schemes
which appear to be "more natural" always give rise to integer shifts.}.
While the essential point stressed, for instance, in \cite{GMM90}\footnote{%
The point of view followed in \cite{GMM90} can be provided with
mathematically sound basis: see \cite{Tu94} and references therein. Similar
results have been obtained in \cite{BN}.} is that in order to agree with the
so called exact skein relations (see for instance \cite{Tu94}) known in knot
theory the parameter $q$ has to depend on $k$ as in Eq. (\ref{ske1}) so that
the shift in Eq. (\ref{ske2}) should refer to the bare unphysical coupling
constant:%
\begin{equation}
k_{bare}\rightarrow k_{bare}+N.  \label{ske3}
\end{equation}%
The presence of Gribov copies can shed some light on the relations between
these two points of view. A first consideration is that in the Abelian 3D
Chern-Simons theory (which has no Gribov ambiguity) there is no such a
shift. Non-trivial Jacobians (which are the source of the shift \cite{EMS89}%
) may arise because of the fact that $\left( \mathcal{A}/G\right) \times G$
is not diffeomorphic\footnote{%
Note that the decomposition $\left( \mathcal{A}/G\right) \times G$\ should
be a global reazlization of the local Faddev-Popov procedure to fix the
gauge in order to take care of the path integral over the gauge degrees of
freedom. In the presence of Gribov ambiguity this local procedure fails to
extend globally.} to $\mathcal{A}$\ (where $\mathcal{A}$\ is the
Chern-Simons configuration space and $G$ is the non-Abelian gauge group):
this phenomenon generates the Gribov ambiguity as well \cite{Si78}. To have
a "pictorial" idea of how Gribov copies manifest themselves one can observe
that in the path integral 
\begin{equation*}
Z(k_{bare})=\int DAD\lambda DcD\overline{c}\exp i\left( k_{bare}S_{0}\left(
A\right) +S_{gf}\left( A,c,\overline{c}\right) \right) ,
\end{equation*}%
the parts affected by the Gribov copies are the gauge fixing and ghosts
parts while, being $k_{bare}$ an integer, the factor $\exp i\left(
k_{bare}S_{0}\left( A\right) \right) $ is constant on the gauge orbit of $A$
(it is worth to recall here that in the present case there is only the gauge
orbit of the vacuum). Due to the "multi-unitons" solutions (see Eq. (\ref%
{mus1})) of Eq. (\ref{cgf2}) one can imagine the following formal splitting
inside the path integral%
\begin{equation}
\int DA\delta \left( \partial _{i}\left( A^{i}\right) \right) \approx \int
DA\dprod\limits_{n=0}^{u}\delta \left( A-A_{n}\right)   \label{fact1}
\end{equation}%
where the $\delta $ of the gauge fixing condition appears after the integral
over $\lambda $. The above product is over the unitons solutions\footnote{$%
n=0$ corresponding to the trivial solution $U=U_{0}$ a constant matrix in
the group which corresponds to a vanishing gauge potential.} $U_{n}$\ of Eq.
(\ref{cgf2}) fulfilling the boundary conditions in Eq. (\ref{bc1})%
\begin{equation*}
A_{n}=U_{n}^{-1}dU_{n}.
\end{equation*}%
The above splitting gives rise to $u+1$ factors 
\begin{equation}
\int DA\dprod\limits_{n=0}^{u}\delta \left( A-A_{n}\right) \exp i\left(
k_{bare}S_{0}\left( A\right) \right) \int DcD\overline{c}\exp i\left(
S_{gf}\left( A,c,\overline{c}\right) \right) .  \label{fact2}
\end{equation}%
This suggests the shift of the coupling constant 
\begin{equation*}
k_{bare}\rightarrow k_{bare}+u+1=k_{bare}+N=k_{ren}
\end{equation*}%
since, as it has been argued in \cite{Wit89}, the shift arises from the
ghosts and gauge fixing terms while $\exp i\left( k_{bare}S_{0}\left(
A\right) \right) $ is constant on the gauge orbit. Of course, the above
formal manipulations are far from being conclusive\footnote{%
In any case, path integral calculations can be made often rigorous in the 3D
Chern-Simons case (see, for instance, \cite{Tu94}).} since the issues of
regularization are rather subtle in Chern-Simons theory. Nevertheless, they
show that the presence of Gribov copies affects the final result and that if
one would restrict the measure to an ambiguity-free region the final result
would be different. Further obstacles to provide the above arguments with a
rigorous basis come from the lacking of a complete understanding of harmonic
maps. Firstly, to the best of the author's knowledge, there is no general
theorem stating that the unitons number $u$ is equal to $N-1$. Furthermore,
there are no complete results in the theory of harmonic map in the cases in
which the spatial topology is more complicated. It is also not perfectly
known the relation between the unitons number and the Casimirs of generic
non-Abelian gauge groups. A mathematical refinement of the above argument
could clarify the non-perturbative origin of the shift in a regularization
independent way. A more "physical" approach to investigate the consequences
of the presence of Gribov copies could be the computation of the
Chern-Simons path integral with the restriction to the modular region.

\section{Conclusions and perspectives}

In the present paper, using powerful tools of harmonic maps and integrable
systems, \textit{all} the Gribov copies in the Coulomb gauge in 3D
Chern-Simons theory have been constructed. This is the first example of a
simple yet non-trivial gauge theory in which all the Gribov copies in the
Coulomb gauge can be determined. The same construction also works in the
case of \textit{all} the Gribov copies of the vacuum in 3D QCD in the
Coulomb gauge. This result gives rise to the possibility to relate the
presence of Gribov copies and the famous shift of the Chern-Simons coupling
constant. It would be very interesting to provide this relation with more
sound basis. It is would be also interesting to compute the Chern-Simons
path integral implementing the restriction to a copies-free region: this
analysis could shed new light (in a case which is much simpler than 4D QCD)
both on the differences between the Gribov and the modular regions and on
whether or not one has to restrict the path integral to an ambiguity free
region.

\section*{Acknowledgements}

I would like to thank A. Anabalon and J. Zanelli: without their questions,
interesting discussions and suggestions this work would not have been done.
I would also like to thank S. Sorella for encouraging comments and important
suggestions and G. Vilasi for discussions and suggestions about integrable
systems. This work was supported by Fondecyt grant 3070055. The Centro de
Estudios Cientficos (CECS) is funded by the Chilean Government through the
Millennium Science Initiative and the Centers of Excellence Base Financing
Program of Conicyt. CECS is also supported by a group of private companies
which at present includes Antofagasta Minerals, Arauco, Empresas CMPC,
Indura, Naviera Ultragas and Telef$o^{\prime }$nica del Sur.

\end{document}